\documentclass[prl,twocolumn,preprintnumbers,amsmath,amssymb,showpacs]{revtex4}
\usepackage{mathrsfs}
\usepackage{bm}
\usepackage[dvips]{epsfig}
\usepackage{bbm}
\usepackage{amsfonts}
\usepackage{color}
\def\bm#1{\mbox{\boldmath{$#1$}}}
\def\rr#1{(\ref{#1})}

\newcommand{\be}{\begin{equation}}
\newcommand{\ee}{\end{equation}}

\begin{document}


\title{Electromagnetic propulsion  and separation by chirality of nanoparticles in liquids}
\author{E. Kirkinis}
\email{kirkinis@amath.washington.edu}
\affiliation{ University of Washington, Department of Applied Mathematics,
Seattle WA 98195}
\author{A.V. Andreev}
 \affiliation{ University of Washington, Department of Physics,
Seattle, WA 98195}
\author{B. Spivak}
 \affiliation{ University of Washington, Department of Physics,
Seattle, WA 98195}
\begin{abstract}
We introduce a new mechanism for the propulsion and separation by chirality of small ferromagnetic particles suspended in a liquid. Under the action of a uniform d.c.
magnetic field $\mathbf{H}$ and an a.c. electric field $\mathbf{E}$ isomers with opposite chirality move in opposite directions. Such a mechanism could have a significant impact on a wide range of emerging technologies
\cite{Lee2007}. The component of the chiral velocity that is odd in ${\bf H}$ is found to be proportional to the intrinsic orbital and spin angular momentum of the magnetized electrons. This effect arises because a ferromagnetic particle
responds to the applied torque as a small gyroscope.
\end{abstract}



\maketitle

Recent years have witnessed an explosion of interest in the fabrication
of nanoscale objects
\cite{Zerrouki2008,Ghosh2009,Zhang2009,Sakar2010}
and their propulsion in liquids.
Numerous mechanisms have been proposed
to achieve this propulsion ranging
from electrophoresis for platinum rods \cite{Paxton2004},
to beating of flexible magnetic rods resembling flagella
\cite{Dreyfus2005}. Some mechanisms
take advantage of the lack of center-symmetry of the
particles and propel them under the action of rotating external fields~\cite{Baranova1978,Ghosh2009,Cheang2010}.

In this article we introduce a new mechanism for the propulsion
and separation by chirality of small ferromagnetic isomers, which are suspended in a liquid and endowed with a
frozen-in magnetic moment $\mathbf{M}$ and electric dipole moment $\mathbf{d}$.
The separation by chirality is induced by
applying a uniform  linearly polarized a.c. electric field $\mathbf{E}$  and a d.c. magnetic field $\mathbf{H}$.  We note that in a given particle the direction of the electric dipole moment ${\bf d}$ is unique. In contrast, the magnetic moment ${\bf M}$ can acquire different orientations with respect to
the crystalline axes.  Here we consider  particles with an easy axis magnetic anisotropy in which states with the two possible magnetization values, ${\bf M}$ and ${\bf -M}$, are realized with equal probabilities. We consider the general case where ${\bf d}$ and ${\bf M}$ are non collinear, and so are ${\bf E}$ and ${\bf H}$. In what follows, we will be interested in the chiral velocity ${\bf V}_{ch}=V_{\circlearrowleft}-V_{\circlearrowright}$
of the particles averaged over different directions of the magnetization.
Here $V_{\circlearrowleft}$ and $V_{\circlearrowright}$ are the velocities of
the right and left-handed particles, respectively.

Since $\mathbf{H}$ is an axial vector and $\mathbf{E}$ a polar vector
the time-averaged chiral velocity of the particle must have the following form
\begin{equation} \label{symmetry}
\mathbf{V}_{ch}= \sigma_{1}\mathbf{h} + \sigma_{2}
(\mathbf{e}\cdot\mathbf{h})\,\mathbf{e}+\sigma_{3} {\bf (e\cdot h)[e\times h]}.
\end{equation}
Here $\mathbf{h}$ and $\mathbf{e}$ are unit vectors in the direction of the
magnetic and electric fields respectively.
The coefficients $\sigma_i$ have opposite signs for isomers of opposite chirality. Thus, particles of opposite chirality will move in opposite directions.

It is important to note that a ferromagnetic particle possesses an intrinsic angular momentum associated with the magnetization, $\mathbf{L}_e= \mathbf{M}/\gamma$~\cite{Landau1958}. The gyromagnetic ratio $\gamma$ can be estimated as $\gamma \sim e/mc$, where  $m$ and $e$ are the electron mass and charge, and $c$ is the speed of light. The value of ${\bf L}_{e}$ is relatively small, and it is usually neglected in studies of the dynamics of small ferromagnetic particles. However the existence of $\mathbf{L}_e \neq \mathbf{0}$ means that the particle responds to external torques as a small gyroscope. This leads to new physical effects.

We show below that the first two terms in Eq.~\rr{symmetry} are proportional to $L_e \equiv |\mathbf{L}_e|$, (the third term, on the other hand, remains finite even as $ L_{e} \to 0$).
This can be seen from the following consideration. In the approximation where $ L_{e}=0$ the magnetic moment affects the particle motion only via the external torque
\begin{equation}
\bm{\tau} =\mathbf{M}\times \mathbf{H}
+\mathbf{d}\times \mathbf{E}.
\label{turk}
\end{equation}
Thus, in this approximation all terms
in the velocity of a particle that are odd in ${\bf H}$ should also be odd
with respect to ${\bf M}$. Consequently, these terms vanish upon averaging over different realizations of ${\bf M}$.
On the other hand, since $\mathbf{L}_e \propto \mathbf{M}$ the chiral current can contain terms which are odd in $\mathbf{M}$ and linear in $\mathbf{L}_e$, which do not vanish upon averaging over different realizations of the particle magnetization $\mathbf{M}$.

The motion of small particles in a dilute suspension can be described using
the formalism of low Reynolds number hydrodynamics~\cite{Happel1965} in which the
external forces, $\bm{F}$, and torques, $\bm{\tau}$, are linearly related to the linear and angular velocities,  $\mathbf{v}$ and $\bm{\omega}$,  by a resistance matrix
\be \label{resistance1}
\left(\begin{array}{c}
\bm{F} \\
\bm{\tau}
\end{array}\right) = \eta \left(
\begin{array}{cc}
\hat{{K}}& \hat{{C}} \\
\hat{{C}} & \hat{{\Omega}}
\end{array}
\right)
 \left(\begin{array}{c}
\mathbf{v} \\
\bm{\omega}
\end{array}\right)
+
 \left(\begin{array}{c}
0 \\
\bm{\omega} \times \mathbf{L}_e
\end{array}\right)
.
\ee
Here $\eta$ is the liquid viscosity. For a particle of characteristic size $R$
the translation $\hat{{K}}\sim R$,
coupling $\hat{{C}}\sim R^2$ and
rotation $\hat{{\Omega}}\sim R^3$ matrices
depend only on the particle's geometry.
Above, we have tacitly assumed that the resistance matrix
is expressed with respect
to a unique fixed point on the particle called the reaction center
which requires that $\hat{C}$ is symmetric. The third
term in Eq.~\rr{resistance1} describes the gyroscopic effect.

We assume that the particles are uncharged so that the electric field does not exert a force on them, $\bm{F}=\bm{0}$. In this case  Eq. \rr{resistance1} gives a linear relation between the propulsion velocity $\mathbf{v}$ and $\bm{\omega}$
\be \label{bodyvelocity}
\mathbf{v}= -\hat{{K}}^{-1} \hat{{C}}\bm{\omega}.
\ee
This relation expresses the so-called propeller effect. Rotation of chiral particles caused by the external torques is accompanied by translational motion. Since the coupling matrix $\hat{C}$ has opposite sign for particles with opposite chirality (while $\hat{K}$ and $\hat{\Omega}$ remain the same), it is clear that  particles of opposite chirality subjected to the same torque will move in opposite directions.

To describe the rotation of the particle we use the body axes defined by the unit vectors $\hat{\mathbf{x}}_1,\hat{\mathbf{x}}_2,\hat{\mathbf{x}}_3$,
whose orientation with respect to the laboratory frame axes $\hat{\mathbf{x}},\hat{\mathbf{y}},\hat{\mathbf{z}}$
is specified by the three Euler angles
$\phi,\theta$ and $\psi$~\cite{Landau1960a}.
Their evolution equation follows from the balance of angular momentum \rr{resistance1},
\be  \label{euler2}
\left(\begin{array}{c}
\dot{\phi} \\
\dot{\theta}\\
\dot{\psi}
\end{array}\right)=
\hat{Q}\widetilde{\Omega}^{-1}_e
\left(\begin{array}{c}
{\tau_{x_1}} \\
{\tau_{x_2}}\\
{\tau_{x_3}}
\end{array}\right).
\ee
Here
\be
\hat{Q} = \frac{1}{\sin\theta}\left(
\begin{array}{ccc}
\sin\psi & \cos\psi & 0\\
\sin\theta \cos\psi& -\sin\theta \sin\psi &0 \\
-\cos\theta \sin\psi& -\cos\theta\cos\psi & \sin\theta
\end{array}
\right)
\ee
is the matrix connecting the particle angular velocity
with the derivatives of the Euler angles~\cite{Landau1960a},
 and
\be
\label{eq:Omega_e}
(\widetilde{{\Omega}}_e)_{ij} = (\widetilde{{\Omega}})_{ij} +
\varepsilon_{ijk}(\mathbf{L}_e)_k,
\ee
is the rotational resistance matrix
 $\widetilde{{\Omega}}
= \eta(\hat{\Omega}-\hat{C}\hat{K}^{-1}\hat{C})$,
 augmented to account for the presence of the intrinsic angular momentum $\mathbf{L}_e$.

In general the torques and forces in Eq. (3) consist of the deterministic torques of Eq. (2) and random torques and forces arising from thermal fluctuations.
In this case the system can be described by the Fokker-Planck equation for the particle distribution function $f=f(t,\mathbf{o})$~\cite{Makino2004}
\be \label{FP4}
\left(\frac{\partial }{\partial t} -
\bm{\mathcal{R}}kT {\widetilde{\Omega}}^{-1}_e\bm{\bm{\mathcal{R}}}\right)f
= \widetilde{\Omega}^{-1}\bm{\mathcal{R}}f\bm{\mathcal{R}}U,
\ee
and the following expression for the ensemble averaged chiral velocity
\be
 \mathbf{V}_{ch} = \nu\int_0^{1/\nu} \! \! \! dt \!\!\int \!\! d^3\mathbf{o}\; \hat{b} \left(kT \bm{\mathcal{R}}f+ f\bm{\mathcal{R}}U  \right).
\label{VFP4}
\ee
Here $T$ is the temperature, $k$ is the Boltzmann constant, $U = - \mathbf{d}\cdot\mathbf{E} -\mathbf{M}\cdot\mathbf{H}$,
$\hat{b} = -\frac{1}{\eta}\hat{\Omega}^{-1}\hat{C}(\hat{K}-\hat{C}\hat{\Omega}^{-1}\hat{C})^{-1}$, $\mathbf{o}$ denotes the particle orientation (specified by the Euler angles),
and $\bm{\mathcal{R}}$ is a vector derivative operator \cite{Landau1958,Favro1960}, whose representation in the body
frame is
\begin{equation}\label{eq:L_definition}
    \left(
      \begin{array}{c}
        \mathcal{R}_{x_1} \\
        \mathcal{R}_{x_2} \\
        \mathcal{R}_{x_3} \\
      \end{array}
    \right)= \left(
               \begin{array}{ccc}
                 \frac{\sin \psi}{ \sin\theta} & \cos \psi & -\cot\theta \sin \psi \\
                 \frac{\cos \psi}{ \sin\theta} & - \sin \psi & -\cot\theta \cos \psi \\
                 0 & 0 & 1 \\
               \end{array}
             \right)\left(
                      \begin{array}{c}
                        \frac{\partial}{\partial  \phi} \\
                        \frac{\partial}{\partial  \theta} \\
                        \frac{\partial}{\partial  \psi} \\
                      \end{array}
                    \right) . \nonumber
\end{equation}
These equations describe diffusion in the particle orientation space and the translational motion associated with its rotation relative to the liquid (we assume that the spatial distribution of particles is uniform).

The strength of thermal fluctuations is characterized by the dimensionless parameters $MH/kT$, and  $dE/kT$.  We consider an electric field of frequency $\nu$, $\mathbf{E}(t)=\mathbf{E} \sin (2\pi \nu t)$. In this case another important parameter is the ratio of the frequency $\nu$ to the rotational equilibration rate,
$\nu\eta R^3/kT$.

For $MH/kT, dE/kT\ll 1$  Eqs.~(\ref{FP4}-\ref{VFP4}) can be solved by perturbation theory. At low frequencies, $\nu\eta R^3/kT \ll 1$,
and to second order in ${dE}$ and first order in $MH$ we obtain
\begin{equation}\label{eq:estimate_fluct}
\sigma_{1}\sim \sigma_{2}\sim
\chi R \nu \frac{ {L}_{e}\nu}{kT}
\left(\frac{dE}{kT}\right)^2\frac{MH}{kT},
\end{equation}
where $\chi \sim K^{-1} C/R$ is a dimensionless measure of the particle chirality. Note that $\sigma_3$ arises only at fourth order in the perturbation theory,
leading to the estimate $\sigma_3 \sim \chi \nu R \nu\eta R^3 (dEMH)^2/(kT)^5$.

In the opposite regime, $MH/kT,dE/kT,\nu\eta R^3/kT \gg 1$,
thermal fluctuations may be neglected and the motion of particles is described by the deterministic equations (\ref{turk}), (\ref{resistance1}) and \rr{bodyvelocity}. In this case, under the influence of $\mathbf{E}$ and $\mathbf{H}$  the particle orientation changes periodically with time and traces a cycle in the space of orientations. Relation \rr{bodyvelocity} shows that propulsion is possible only if this cycle is non-self-retracing. This is the analogue of the ``clam shell'' theorem~\cite{Purcell1977} for forced propulsion at low Reynolds numbers. It is worth noting that although the cycle traced by the linearly polarized electric field is self retracing, the cycle traced by body orientations is not.

At low frequencies where $\nu \ll \frac{dE}{\eta R^3}$ and $\nu \ll  (MH)^2/(d E \eta R^3)$,
the frequency dependence of the
propulsion velocity is linear
\be
\label{Vch}
\sigma_{1,2}  \sim \chi \nu R  (L_{e}/\eta R^{3}) ,
\ee
in contrast to the quadratic dependence
in the strong fluctuation regime (\ref{eq:estimate_fluct}).
The $\sigma_3$ component of velocity has a superlinear dependence on the frequency, see Eq.~(\ref{eq:sigma_3}) below. Therefore at low frequencies the propulsion velocity is confined to the $\mathbf{E}$-$\mathbf{H}$ plane.

The linear frequency dependence of the propulsion velocity is a feature which often arises in the adiabatic regime, where the
orientation of the particle corresponds to instantaneous equilibrium. This occurs, for example,  in the case where the electric field $\mathbf{E}$ is circularly polarized~\cite{Baranova1978,Ghosh2009}.

For a linearly polarized electric field and constant magnetic field, the origin of the linear dependence on $\nu$ in Eq.~(\ref{Vch}) is more subtle. In this case the adiabatic approximation holds during most of the oscillation period. During these intervals the particle orientation is determined by the instantaneous values of ${\bf E}(t)$ and $\mathbf{H}$, and the trajectory of the particle is self-retracing. However, the adiabatic approximation is violated
at time intervals where ${\bf E}(t)$ is approximately zero.
During these times the equilibrium orientation of the particle is not unique: the particle can rotate freely about the axis pointing along $\mathbf{M}$, which is aligned with $\mathbf{H}$. Most of the particle's rotation and propulsion occurs near these instances. As the electric field changes sign the particle rotates
about $\mathbf{M}$ by an angle $\pi$. The direction of rotation is determined by corrections to the adiabatic approximation, which break the symmetry between clockwise and anticlockwise rotations,   and lead to a non-self retracing cycle and non-vanishing propulsion velocity linear in $\nu$. However, if $\mathbf{L}_{e}=0$, upon reversal $\mathbf{M}\to -\mathbf{M}$, the direction of the rotation and propulsion velocity also reverses. Therefore $\mathbf{V}_{ch}$ averaged over realizations of $\mathbf{M}$ vanishes. Accounting for the finite value of $L_e$ results in the difference of the resistance tensors $\widetilde{\Omega}_e$ in states with $\pm\mathbf{M}$, and incomplete cancellation of contributions to ${\bf V}_{ch}$ from particles with different values of ${\bf M}$.  This leads to Eq.~(\ref{Vch}).

\begin{figure}[h]
\begin{center}
$$ \hbox{\psfig{file=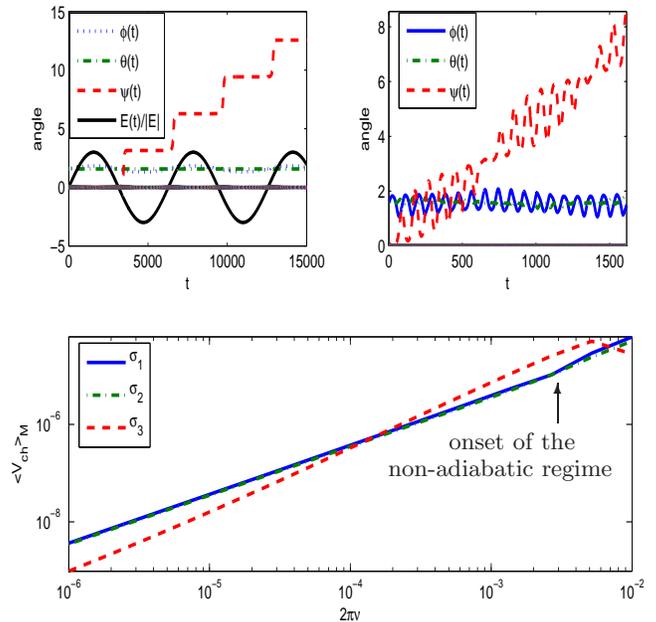,
height=3.6in,width=3.8in,angle=0}
\put(-95,75){onset of the}
\put(-118,65){non-adiabatic regime}
\put(-54,85){\vector(0,1){15}} 
}
$$ \caption{(color online) Upper left: time dependence of the electric field and the Euler angles at low frequencies. The angle $\psi$ increases in a stepwise fashion.
Upper right: breakdown of the
staircase at higher frequencies.
Lower: log log plot of the averaged-over-M
chiral velocity versus frequency. The frequency exponent of $\sigma_1$ and
$\sigma_2$ (overlapping straight lines) is $1$ at low frequencies and
that of $\sigma_3$ is (approximately) $5/4$ as discussed above Eq.~\rr{eq:sigma_3}.
For higher frequencies the motion enters a non-adiabatic regime which
changes the values of the aforementioned exponents, a fact
reflected by
cusp-like features in the frequency dependence of the propulsion velocity.
\label{doubleangle} }
\end{center}
\end{figure}

\begin{figure}
\begin{center}
$$ \hbox{\psfig{file=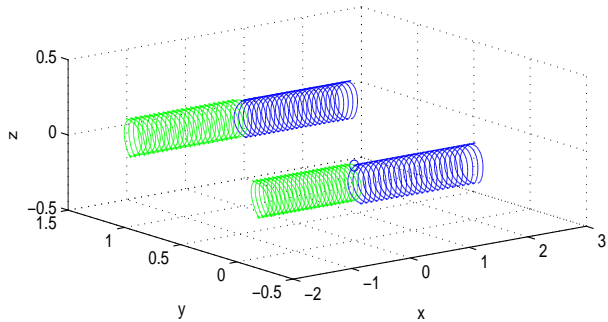,
height=1.8in,width=3.5in,angle=0}}
$$ \caption{(color online) Particle trajectories in the laboratory frame for $\mathbf{H}$ and $\mathbf{E}$ directed along the $x$ and $y$ axes respectively.
The motion of particles with opposite values of $\mathbf{M}$ are shown in blue and green. The trajectories originate from the same point and move on average in the positive (blue) and negative (green) $x$-direction.
The trajectories in the background are calculated with $L_e=0$. In this case the displacements for opposite $\mathbf{M}$ are opposite. The  trajectories in the foreground are computed with $L_e \neq 0$. In this case the magnitudes of displacements for opposite values of $\mathbf{M}$ are different.
\label{nature_lab3} }
\end{center}
\end{figure}

Numerical solutions of the equations of motion, Eq.~(\ref{resistance1}) confirm this picture. We choose $\hat{x}_3 \parallel \mathbf{M}$, so that the Euler angle $\psi(t)$ corresponds to rotation of the particle about $\mathbf{M}$. Typical results of numerical solutions
of Eq.~\rr{resistance1} are presented in Figures
\ref{doubleangle} and \ref{nature_lab3}.
The time evolution of the angle
$\psi$ exhibits a staircase structure, increasing (on average) linearly with
$t$ (see the upper panels in Fig.\ref{doubleangle}). At low frequencies there are two steps per oscillation period and each step corresponds to a rotation of the particle about $\mathbf{M}$ by an angle $\pi$,
as in the upper left of Fig.\ref{doubleangle}. In this regime the $\sigma_1$ and $\sigma_2$ components of the propulsion velocity are linear with respect to the
frequency $\nu$, whereas the $\sigma_3$ component scales as $\nu^{1+\alpha}$ with $\alpha\approx 0.25$, as shown in the lower panel in
Fig.\ref{doubleangle}.  Propulsion occurs near the instances $t_n=n/\nu$, where the electric field changes sign. In these intervals the electric field changes linearly with time, $E(t) \approx  2\pi \nu E (t-t_n)$. Therefore deviations from adiabaticity depend only on the product $\nu E$. This gives the following estimate for $\sigma_3$ in the low frequency regime,
\begin{equation}\label{eq:sigma_3}
  \sigma_{3}\sim \chi \nu R
\left[\frac{\nu \eta R^3 dE}{(MH)^2}\right]^{\alpha}.
\end{equation}
As the frequency $\nu$ increases, the character of the motion undergoes a series of bifurcations: the step-like character of evolution of $\psi(t)$ is preserved but the steps become separated by several oscillation periods (see the upper right panel in Fig.\ref{doubleangle}). This leads to cusp-like features in the frequency dependence of the propulsion velocity (see the lower panel in Fig.~\ref{doubleangle}).

When $L_e=0$, particles with opposite values of $\mathbf{M}$ rotate and move in opposite directions in the
$\mathbf{H}$-$\mathbf{E}$ plane, leading to a vanishing planar displacement upon averaging over the directions of $\mathbf{M}$.
In contrast, when $L_e\neq 0$ the in-plane displacement averaged over
$\mathbf{M}$ is finite. This is illustrated in Fig.~\ref{nature_lab3} for the case when $\mathbf{H}$ and $\mathbf{E}$ are perpendicular to each other.

We now estimate the magnitude of the above effect. Assuming that a single domain ferromagnetic particle is roughly spherical, we get an estimate for  the dimensionless parameter characterizing the magnitude of the gyroscope effect,
\be
\label{eq:L_e_estimate}
\frac{L_e}{\widetilde{\Omega}} \sim \frac{s\hbar  n}{6 \eta},
\ee
independent of the particle size. Here $n$ is the volume density of magnetic atoms,  $\hbar$ is Planck's constant, and $s$ is the spin per atom (in units of $\hbar$), which in different materials can lie in the range $1-10^{-2}$.
Using the viscosity of water at normal conditions, $\eta \sim 10^{-2}$g$/$cm$\cdot$s,  and $n\sim 10^{23}$cm$^{-3}$ we get $L_{e}/\widetilde{\Omega} \sim (s/6) \times 10^{-2}$.
The estimate Eq.~(\ref{Vch}) holds provided  the inequalities $\nu < d E /(\eta R^3 )$, $ \nu < (MH)^2 /(\eta R^3 dE)$
and  $dE,MH>kT$ are satisfied. In modern
experiments~\cite{Washizu1990} electric fields in excess of $10^6$ V$/$m at frequencies $1$ MHz have been realized in aqueous solutions. Estimating  $d\sim eR$ we see that the required inequalities are satisfied for a particle size $R \sim 100$ nm. The magnetic restriction, $MH/kT \gg 1$, is satisfied even in weak magnetic fields for a ferromagnetic particle of this size. Assuming that the degree of chirality is $\chi \sim 0.1$,
$H\sim 10$ Gauss and $d\sim 10^4 D$ and using Eqs.~(\ref{Vch}) and (\ref{eq:sigma_3}) with the aforementioned electric
fields and frequencies we get the estimates
\begin{equation}
\sigma_{1,2}\sim  (0.1 - 10)\,  \mu\text{m}/\text{s},  \,\,\,\
\sigma_{3}\sim 1 \, \text{mm}/\text{s},
\end{equation}
which show that the effect is detectable.

Finally, we point out the existence of
another class of effects that are generically related
to the one discussed above.
These are realized by replacing the linearly polarized a.c.
electric field ${\bf E}$ with either a gradient of temperature $\nabla T$,
pressure $\nabla P$, or an oscillating
magnetic field $\tilde{{\bf H}}=\tilde{H}\tilde{{\bf h}}$, where $\tilde{H}=a \sin (2\pi \nu t)$.
To describe these effects on the phenomenological level one should make the following changes in Eq. \rr{symmetry}:
${\bf e}\rightarrow \nabla T$, ${\bf e}\rightarrow \nabla P$, and  ${\bf e}\rightarrow \tilde{{\bf h}}$ respectively. Calculating the corresponding
coefficients which are analogues of the $\sigma_{1,2,3}$ in Eq.~\rr{symmetry}
is beyond the scope of the present article.

\noindent
\textbf{Acknowledgments}: We are grateful to E. Ivchenko for useful
discussions. This work was supported by  DOE grant DE-FG02-07ER46452 (EK and
AVA) and NSF grant DMR-0704151 (BS).

\end{document}